# Broadband energy-efficient optical modulation by hybrid integration of silicon nanophotonics and organic electro-optic polymer

Xingyu Zhang*[,a], Amir Hosseini[b], Harish Subbaraman[b], Jingdong Luo[c], Alex K.-Y Jen[c],
Chi-jui Chung[a], Hai Yan[a], Zeyu Pan[a], Robert L. Nelson[d], and Ray T. Chen*[,a,b]

[a] University of Texas at Austin, 10100 Burnet Rd, MER 160, Austin, TX 78758, USA;
[b] Omega Optics, Inc., 8500 Shoal Creek Blvd, Austin, TX 78757, USA;
[c] University of Washington, 302 Roberts Hall, Seattle, Washington 98195, USA;
[d] Air Force Research Laboratory at Wright Patterson, Dayton, Ohio 45433, USA.

## ABSTRACT

Silicon-organic hybrid integrated devices have emerging applications ranging from high-speed optical interconnects to photonic electromagnetic-field sensors. Silicon slot photonic crystal waveguides (PCWs) filled with electro-optic (EO) polymers combine the slow-light effect in PCWs with the high polarizability of EO polymers, which promises the realization of high-performance optical modulators. In this paper, a broadband, power-efficient, low-dispersion, and compact optical modulator based on an EO polymer filled silicon slot PCW is presented. A small voltage-length product of $V_\pi \times L=0.282 V \times mm$ is achieved, corresponding to an unprecedented record-high effective in-device EO coefficient ($r_{33}$) of 1230pm/V. Assisted by a backside gate voltage, the modulation response up to 50GHz is observed, with a 3-dB bandwidth of 15GHz, and the estimated energy consumption is 94.4fJ/bit at 10Gbit/s. Furthermore, lattice-shifted PCWs are utilized to enhance the optical bandwidth by a factor of ~10X over other modulators based on non-band-engineered PCWs and ring-resonators.

## 1. INTRODUCTION

The combination of silicon photonics and electro-optic (EO) polymers has enabled compact and high-performance hybrid integrated microwave photonic devices [1], such as modulators [2], interconnects [3] and sensors [4]. The large EO coefficient ($r_{33}$), ultrafast response time, very low dispersion, and spin-coating feature of EO polymers promise low-power consumption, ultra-high speed operation, and ease of fabrication [5-8]. Silicon photonics offers the potential of complementary metal–oxide–semiconductor (CMOS) compatible photonic integrated circuits [9, 10]. Silicon photonic crystal waveguides (PCWs) [11, 12] exhibit slow-light effects which are beneficial for device miniaturization [13, 14]. Especially, EO polymer filled silicon slotted PCWs [15] further reduce the device size and enhance the device performance by combining the best of these two systems.

In this paper, we report the design, fabrication, and characterization results of a high-performance silicon-organic hybrid (SOH) PCW modulator with high modulation efficiency, broad RF bandwidth, low power consumption, and low optical dispersion. This paper is organized as follows. The design of a band-engineered PCW modulator and the principle of high-speed modulation are presented in the section on "principle and design". The section on "fabrication" describes the process of device fabrication. In the section on "results and discussion", we report the characterization results of our modulator, including $V_\pi \times L$, RF bandwidth, power consumption, and optical bandwidth. In the "conclusion" section, summary and future work are presented.

## 2. DEVICE DESIGN

*xzhang@utexas.edu; phone 1 512-471-4349; fax 1 512 471-8575
*raychen@uts.cc.utexas.edu; phone 1 512-471-7035; fax 1 512 471-8575

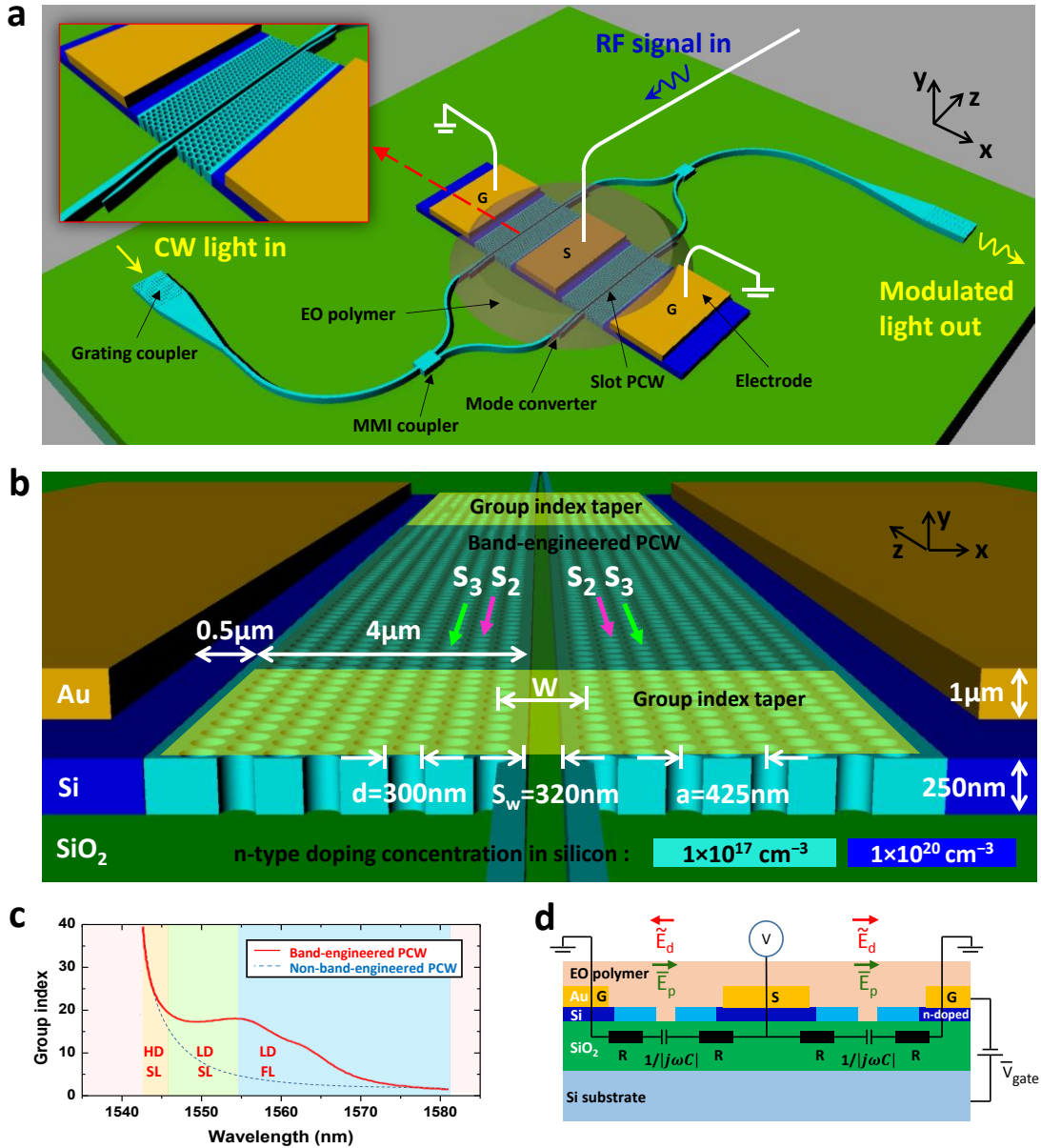

Fig. 1. **An EO polymer filled silicon slot PCW MZI modulator designed on an SOI substrate. a**, Three-dimensional schematic of the modulator. The inset shows the magnified image of the silicon slot PCW on one arm of the MZI. PCW: photonic crystal waveguide; MMI: Multi-mode interference; G: ground electrode; S: signal electrode. **b**, A tilted view of the slot PCW on one arm of the MZI, showing the cross-sectional device dimension, 2-level doping concentrations, group index taper region, and band-engineered PCW region. Note: the EO polymer is not shown here for better visualization. **c.** Simulation result of engineered group index in the slot PCW (red curve) as a function of wavelength, showing 8nm low-dispersion slow-light wavelength region (flat band nature of low-dispersion region highlighted in green). Also overlaid is a blue dashed curve representing the dispersive group index versus wavelength for non-band-engineered PCW for comparison. HD SL: high-dispersion slow-light; LD SL: low-dispersion slow-light; LD FL: low-dispersion fast-light. **d**, Equivalent electrical circuit of the MZI modulator in a push-pull configuration, with a constant gate voltage applied on the bottom silicon substrate. $E_d$: driving field, $E_p$: poling field, $V_{gate}$: gate voltage.

Our optical modulator is a symmetric Mach-Zehnder Interferometer (MZI), with slot photonic crystal waveguides (PCWs) incorporated in both the arms, as shown in Fig. 1a. We start with a silicon-on-insulator (SOI) substrate with 250nm-thick top silicon and 3μm-thick buried oxide (BOX) layers. The slot and holes of the PCWs are filled with an EO polymer (SEO125 from Soluxra, LLC), which has a refractive index, n=1.63 at 1550nm, and an exceptional combination of large EO coefficient ($r_{33}$ of ~100pm/V at 1550nm), low optical loss, synthetic scalability, as well as excellent photochemical stability. Its relatively high glass transition temperature of 150℃ provides good temporal stability, and the EO coefficient of poled SEO125 is essentially unchanged under ambient conditions. The refractive index of the EO polymer can be changed by applying an electric field via the Pockels effect, and is given as $\Delta n = -\frac{1}{2} r_{33} n^3 V / S_w$, where Δn is the change in refractive index of the EO polymer, V is the applied voltage, $S_w$ is the slot width. The slot PCW has a hexagonal lattice of air holes with the lattice constant a=425nm, hole diameter d=300nm, slot width $S_w$=320nm, and center-to-center distance between two rows adjacent to the slot W=1.54($\sqrt{3}$)a. The optimized slot width of 320nm supports a confined optical mode, and also tremendously increases the EO polymer poling efficiency by suppressing the leakage current through the silicon/polymer interface during the poling process [16]. We also note that the poling-induced optical loss is reduced by this reduction of leakage current [17]. More importantly, different from typical slot widths of 100~120nm in conventional slot waveguides [15, 18], widening the slot width to 320nm reduces the slot capacitance, enabling the potential of higher RF bandwidth and lower power consumption, and relaxes the fabrication complexities. To address the issue of the narrow operational optical bandwidth of typical PCW modulators (less than 1nm at the group index, $n_g$>10) [2, 19], lattices of the second and third rows of the PCW are shifted parallel to the slot with relative values of $S_2$ = -85nm, $S_3$ = 85nm (indicated by the arrows in Fig. 1b). As a result, a flat group index ($n_g$) of 20.4 (±10%) over a wavelength range from 1546nm to 1554nm is achieved, as shown in Fig. 1c, enabling an optical spectrum range as wide as 8nm for low-dispersion operation. In order to efficiently couple light from a strip waveguide into and out of the slot PCW, an adiabatic strip-to-slot mode converter is designed [20]. To make a smooth transition between the group indices from a slot waveguide ($n_g$~3) to a slot PCW ($n_g$~20.4), a group index taper consisting of 8 periods of non-lattice-shifted PCW is developed, in which W increases parabolically from W=1.45($\sqrt{3}$)a to W=1.54($\sqrt{3}$)a [21]. Sub-wavelength gratings (SWGs) are used to couple light into and out of the silicon strips [22]. Multi-mode interference (MMI) couplers are used for beam splitting/combining [21]. The PCW interaction length is chosen to be 300μm for $V_\pi$<1V based on theoretical calculation using $L = \frac{1}{2\sigma} \cdot \frac{n}{\Delta n} \cdot \frac{\lambda}{n_g}$ [23], where σ=0.33 is the confinement factor in the slot [24] calculated by the simulation, λ=1550nm is wavelength, and $n_g$=20.4.

Due to the short interaction length (300μm), the maximum modulation frequency of our modulator is not limited by the group velocity mismatch between RF and optical waves, which is usually the case in conventional modulation devices, necessitating the use of complex traveling wave electrode geometries. Instead, it is mainly limited by the time needed to charge the capacitor formed by the slot through the finite ohmic resistance across the silicon; therefore, our modulator can be driven by lumped electrodes [25-27]. The silicon PCW is selectively implanted by n-type dopant (Phosphorus ion, 31P+) with ion concentrations of $1\times10^{20}$cm$^{-3}$ and $1\times10^{17}$cm$^{-3}$ [28], as shown in Figs. 1a and b, so that the resistivity of silicon region is reduced to $9\times10^{-6}$Ω·m and $9\times10^{-4}$Ω·m, respectively [29]. The purpose of using relatively lower concentration ($1\times10^{17}$cm$^{-3}$) in the waveguide region is to avoid significant impurity-induced optical scattering loss [30, 31]. The separation between the gold electrodes is 9.32μm. Figure 1d shows a simplified equivalent circuit of the modulator driven in a push-pull configuration, in which the slot can be represented by a capacitor C and the silicon PCW region by a resistor R. As the modulation frequency increases, the percentage of electric potential dropped across the slot will decrease due to the reduced slot impedance ($1/|j\omega C|$). The low resistivity of doped silicon can help increase the electric field inside the slot at high frequencies. Simulations by COMSOL Multiphysics show that over 90% of the electric potential is dropped across the slot at 10GHz. Both the optical field and the modulation RF field are concentrated in the 320nm-wide slot, enabling a large field interaction factor, and thus providing efficient modulation at high modulation frequency. Based on simulations performed using Lumerical Device software, the total resistance of the 300μm-long silicon PCW is 189 Ohms, and the slot capacitance is as small as 39fF. Thus, the theoretical 3-dB modulation bandwidth of the MZI modulator is estimated to be $1/(2\pi RC)$=22GHz.

It was recently demonstrated that the RF bandwidth of modulators can be further improved by applying a constant gate voltage ($V_{gate}$) between the bottom silicon substrate and top silicon layer [32, 33] to make the top silicon layer sufficiently conductive. This technique avoids the need for heavy doping, so impurity-scattering optical loss can be minimized [34]. This method was used for conventional silicon slot waveguides to achieve modulation up to 42.7GHz [32] and low energy consumption of 320fJ/bit [33]. Here we apply a similar technique on our silicon PCW modulator, as shown in Fig. 1d. By applying a positive voltage on the backside silicon substrate (weakly doped, resistivity of ~15Ω·cm) across the 3μm-thick BOX layer of our device, the energy bands in the n-type top silicon are bent, and thus more electrons accumulate at the interface between the silicon PCW and the BOX layer. Since the resistivity of the silicon region is inversely proportional to the density and mobility of majority free carriers, the resistivity of the silicon PCW region can be reduced, leading to an enhanced RF bandwidth and a reduced power consumption of the modulator.

## 3. FABRICATION

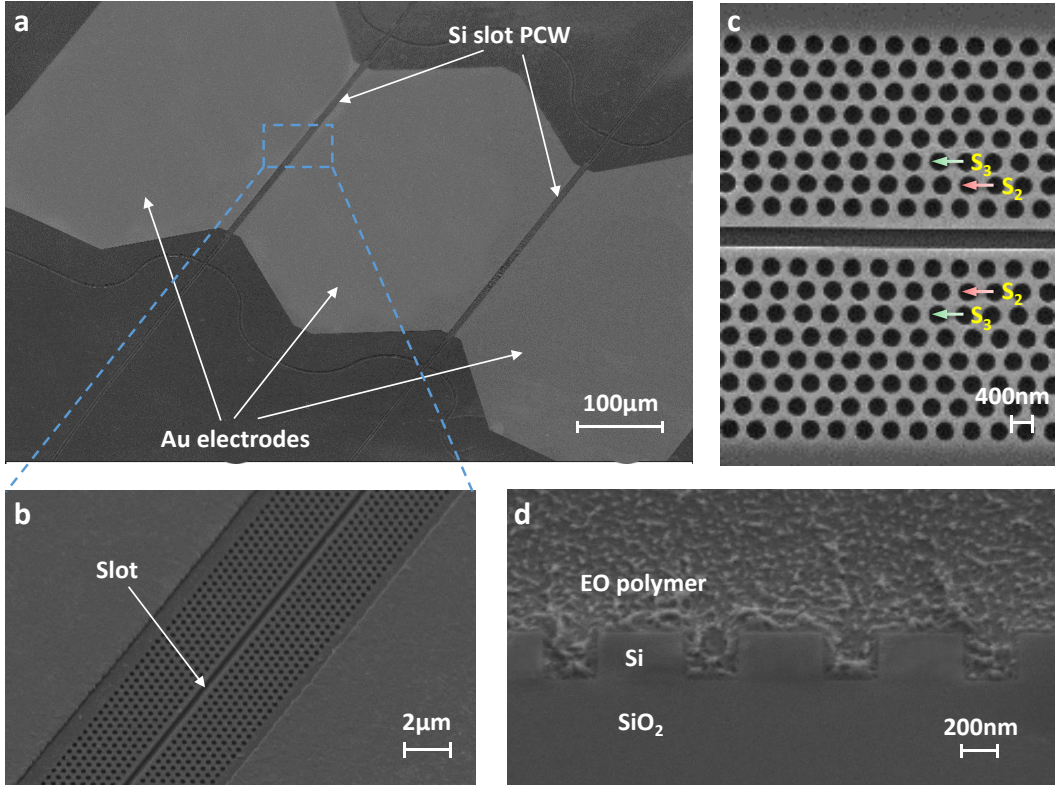

Fig. 2. **SEM images of the fabricated device. a,** A tilted view of the symmetric MZI modulator with silicon slot PCWs in both arms. **b**, A magnified image of the silicon slot PCW in one arm inside the gap of electrodes. **c**, A top view of the slot PCW, with arrows indicating the shifted lattices on the second and third rows. $S_2$=-85nm, and $S_3$=85nm. **d**, A cross-sectional view of the photonic crystal structure filled with EO polymer.

    The fabrication procedure starts with an SOI wafer. The silicon slot PCW is patterned by electron-beam lithography and reactive ion etching (RIE). Then, the silicon slot PCW is first implanted with 31P+ at energy of 92keV and dose of $1.05 \times 10^{12}/cm^2$ to reach an ion concentration of $1 \times 10^{17} cm^{-3}$. Next, the device is patterned by photolithography and selectively implanted with 31P+ at energy of 92keV and dose of $1.05 \times 10^{15}/cm^2$ to reach an ion concentration of $1 \times 10^{20} cm^{-3}$ in the region which will connect the gold electrodes in order to form ohmic contacts. A rapid thermal annealing at 1000°C for 10min in a flowing nitrogen environment is followed to annihilate the induced defects and activate the implanted ions, which also improves the optical performance of the ion-implanted waveguides. Next, 1μm-thick gold electrodes with 5nm-thick chromium adhesion layers are patterned using photolithography, electron-beam evaporation, and lift-off. Figures 2a and b show the SEM images of the fabricated device in a tilted view. Figure 2c shows a top view of the fabricated slot PCW, with arrows indicating the shifted lattices. Next, the EO polymer is formulated, and then covered over the PCW and dispersed into the holes and slot by spin coating. Figure 2d shows the cross section after EO polymer filling.

    Finally, to activate the EO effect, a poling process is performed [35-37]. The device is heated up on a hot plate to the EO polymer glass transition temperature of 150°C in a nitrogen atmosphere, and a constant poling electric field of 110V/μm is applied across the EO polymer inside the slot in a push-pull configuration, as shown in Fig. 1d. The randomly oriented chromophore dipoles inside the polymer matrix are then free to rotate and align in the direction of poling electric field. Next the temperature is quickly decreased to room temperature while the constant electric field is still applied, and eventually the chromophores are locked in a uniform direction to form a noncentrosymmetric structure. During this poling process, the leakage current is monitored and it remains below 0.53nA, corresponding to a low leakage current density of 5.5A/m$^2$. This is comparable to the typical leakage current density of 1-10A/m$^2$ measured in a thin film configuration, indicating a high poling efficiency [38].

## 4. CHARACTERIZATION

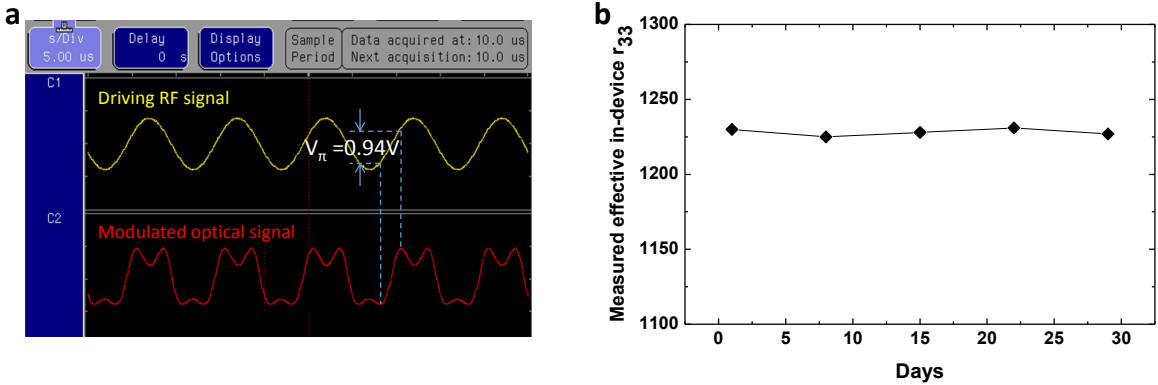

Fig. 3. **Device characterization at low frequency. a,** Transfer function at 100kHz. The $V_\pi$ is measured to be 0.94V from over-modulation. **b,** Measured effective in-device $r_{33}$ as a function of time in days, indicating the long-term stability of the modulator.

A low-frequency modulation test is first performed on the device to measure the voltage-length product, $V_\pi \times L$, which is a figure of merit (FOM) for optical modulators. TE-polarized light from a tunable laser source (1550nm) is coupled into and out of the device utilizing an in-house built grating coupler setup [22]. An RF signal from a function generator is applied onto the electrodes in a push-pull configuration, as shown in Fig. 1d. The modulator is biased at the 3dB point and driven by a sinusoidal RF wave with a peak-to-peak voltage of $V_d$=1.5V at 100KHz. The modulated output optical signal is detected using an amplified avalanche photodetector and a digital oscilloscope setup. As shown in Fig. 3a, over-modulation is observed on the output optical waveform, and the $V_\pi$ of the modulator is measured to be 0.94V. Thus, the FOM of the modulator achieved is $V_\pi \times L$=0.94V×300μm=0.282V×mm. The effective in-device $r_{33}$ is then calculated to be [4]

$$r_{33,eff} = \frac{\lambda S_w}{n^3 V_\pi \sigma L} = 1230 \text{ pm/V} \tag{1}$$

where λ=1550nm, $S_w$=320nm, n=1.63, L=300μm, and σ=0.33, which is the highest ever in-device $r_{33}$ ever recorded. Such a high $r_{33}$ value originates from the combined effects of a large bulk $r_{33}$ of the EO polymer material, an improved poling efficiency achieved via widening the slot width (320nm), the slow-light enhancement in the silicon PCW, as well as the increased percentage of voltage drop across the slot due to silicon doping. Discounting the slow-light effect, the actual in-device $r_{33}$ is estimated to be as high as 98pm/V [39-41]. In addition, to verify the long-term stability of the device, the same test is repeated in the same conditions over the duration of a month, and the measured effective in-device $r_{33}$ as a function of time in days is shown in Fig. 3b. It can be seen that no severe degradation of device performance is observed after a month, due to the improved stability of the EO polymer material.

The RF bandwidth is measured in a small signal modulation test. RF driving signal is provided by a vector network analyzer (VNA) and applied onto the electrodes of the modulator via a ground-signal-ground (GSG) picoprobe. The modulated optical signal is amplified by an erbium doped fiber amplifier (EDFA) and received by a high-speed photodetector, and then the received power is measured using a microwave spectrum analyzer (MSA). The measured EO response of the device as a function of modulation frequency is shown in Fig. 4a, from which a 3-dB bandwidth of 11GHz is measured. Note that the upper frequency of this measurement is limited by the upper limit of our MSA, which is 26GHz.

Next, in order to overcome this measurement limit and demonstrate the modulation response at frequencies over 26GHz, we perform another measurement using a sideband detection technique [42-45]. The optical output of the modulator is directly connected to the optical spectrum analyzer (OSA), and the transmission spectrum of the modulator is measured. When the modulator is driven by a high frequency RF signal, two sidebands appear in the transmission spectrum, equally spaced around the main peak [42-46]. Figure 4b shows overlaid transmission spectra of the optical modulator driven at 10GHz, 20GHz, 30GHz and 40GHz. At higher modulation frequencies, the power of the sidebands becomes lower due to the combined effects of decreased electric potential drop across the slot, reduced output power of the RF source, and increased RF loss on the feeding cable and the probe. Since the power of the main peak and first sideband is proportional to the square of the zero-order and first-order Bessel function of the first kind ($J_i$, i=0,1) as a function of phase modulation index (η) which represents the achieved phase shift (unit:

radians), by measuring the ratio of the main peak power and sideband power ($J_0^2(\eta)/ J_1^2(\eta) \approx (2/\eta)^2$), the phase modulation index ($\eta$) can be extracted [32, 44-46]. The obtained modulation index as a function of modulation frequency is plotted as the red curve shown in Fig. 4c. Sideband signals are observed above the noise floor until the modulation frequency is over 43GHz.

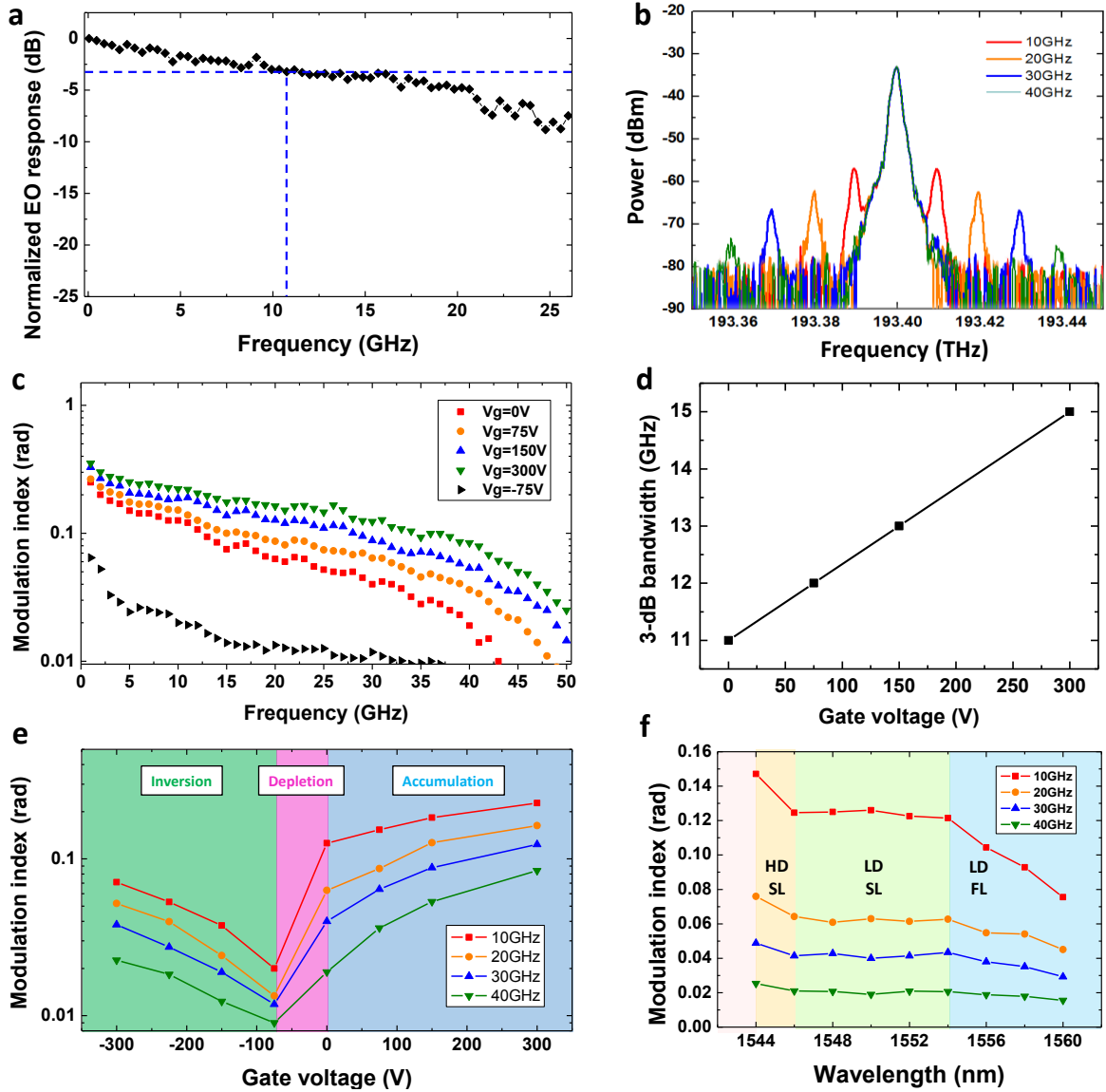

**Fig. 4 Device characterization at high frequency. a,** Measured normalized EO response of the modulator as a function of modulation frequency in a small-signal modulation test. The 3-dB bandwidth is measured to be 11GHz. **b,** Measured optical transmission spectra of the modulator operating at 10GHz, 20GHz, 30GHz and 40GHz. **c,** Measured modulation index as a function of frequency, under different backside gate voltages. **d,** Increased 3-dB RF bandwidth as the positive gate voltage increases. **e,** Measured modulation index as a function of $V_{gate}$ at different modulation frequencies, overlaid with the states of accumulation, depletion and inversion. **f,** Measured modulation index over a range of optical wavelengths. The modulation index is nearly constant over a low-dispersion slow-light region of 8nm.

Next, to further increase the RF bandwidth of the modulator, a positive gate voltage ($V_{gate}$) from a high-voltage supply is applied to the bottom silicon substrate across the BOX layer of our device [32, 33], as shown in Fig. 1d. The positive $V_{gate}$ is varied and the corresponding modulation index is measured. As shown in Fig. 4c, as the positive $V_{gate}$ increases, the measured modulation index at each frequency increases and the whole curve becomes flatter, due to the increased electron accumulation at the interface of the silicon PCW and the BOX [32, 33]. When the $V_{gate}$ is increased over 150V, the sideband power starts to appear above the noise level in the transmission spectrum at 50GHz, indicating the achievable frequency response of our modulator at up to 50GHz. This measured maximum modulation frequency is limited by the available frequency range of our RF source. Figure 4d shows the 3-dB bandwidth of the modulator as a function of positive $V_{gate}$, and it can be seen that under the $V_{gate}$ of 300V, the 3-dB bandwidth is increased to 15GHz. Note the breakdown electric field of the silicon dioxide is about 0.5GV/m [47], corresponding to a voltage of 1500V that the 3 μm-thick BOX layer can withstand.

In order to further investigate the device performance under $V_{gate}$, a negative $V_{gate}$ is applied, and the modulation index is measured and plotted in Fig. 4e. It can be seen that with the magnitude of negative voltage slightly increased, the measured modulation index decreases due to the depletion of electrons. At a $V_{gate}$ of around -75V, the free electrons are almost completely depleted, so the modulation index becomes the smallest, which is also shown by the black curve in Fig. 4c. When the magnitude of the negative voltage further increases, modulation index starts to increase because "inversion" state occurs in which holes are accumulated in the top silicon PCW layer. This interesting phenomenon is quite similar to the well-known Metal-Oxide-Semiconductor (MOS) capacitor structure [48].

What is more, a small switching voltage and low power consumption is achieved under a high positive $V_{gate}$. For example, the measured modulation index is η=0.23 at 10GHz under $V_{gate}$=300V, and correspondingly, the required switching voltage is calculated to be $V_\pi = \pi/\eta \times V_d$ =2.2V at 10GHz, where $V_d$= 0.16V is the RF driving voltage calculated from the output power of RF source. Since our modulator is a lumped device without termination, the power consumption is then dominated by the capacitive load of the slot. The RF power consumption for 100% modulation depth is $2\pi f \times (\tfrac{1}{2} C V_\pi^2) \times 2$ =24mW at modulation frequency of f=10GHz, where C=39fF is the slot capacitance calculated by the simulation, $V_\pi$=2.2V is used as the driving voltage to achieve a maximum extinction ratio, and a factor of 2 is added due to the push-pull configuration. In addition, we make an estimation of energy consumption per bit for our device [49]. If our modulator is driven by PRBS signals with the same power level, we estimate the energy consumption per bit for our modulator at the bit rate of 10Gbit/s as $W_{bit} = \tfrac{1}{4} C V_\pi^2 \times 2$ = 94.4fJ/bit [26, 33, 45, 49, 50]. Note that, in actual high-speed digital modulations, the driving voltage can be smaller than $V_\pi$, in which case a decently clear eye diagram, a high enough extinction ratio and acceptable bit error rate (BER) can be still achieved using lower energy [14, 19, 33, 51, 52]. Though, we still use $V_\pi$ as driving voltage (i.e.100% modulation depth) in our estimation, because this can compensate the actual voltage drop caused by experimental imperfections such as reflections, drift, RF loss, etc., and allows for a reasonable estimated value of energy consumption. This very low estimated energy consumption is due to both a significantly reduced $V_\pi$ and the very small capacitance achieved by widening the slot. Note that, although the applied $V_{gate}$ is high, the power consumption on the backside gate is negligible (<30pW) due to the highly insulating BOX layer.

Finally, to demonstrate the wide optical bandwidth of this PCW modulator, the wavelength of the laser input is tuned from 1544nm to 1560nm, while $V_{gate}$ is set to be zero and all other testing conditions are kept the same. Over this spectrum range, the modulation index is measured at 10GHz, 20GHz, 30GHz, and 40GHz, and the results are plotted in Fig. 4f. It can be seen that at each modulation frequency, the curve of the measured modulation index looks flat from 1546nm to 1554nm, with a small variation of ±3.5%. This is because the modulation index is proportional to the $n_g$ (η~1/$V_\pi$ and $V_\pi$~B×λ/$n_g$, where B is a constant) [53], and $n_g$ has been engineered to be almost constant in this low-dispersion slow-light wavelength region, as shown in Fig. 1c. This 8nm-wide low-dispersion spectrum range is useful for some applications such as wavelength division multiplexing (WDM), and also makes our modulator insensitive to variations of wavelength and temperature, which is much better than non-band-engineered PCWs [2, 19] and ring resonators [54-57]. In Fig. 4f, the largest modulation index is achieved in the high-dispersion slow-light region (from 1543.5nm to 1546 nm), because of the largest $n_g$ in this region. As the wavelength increases over 1554nm, the measured modulation index decreases due to the decreasing value of $n_g$.

## 5. DISCUSSIONS

In recent years, quite a few research groups reported impressive work on analog/digital optical modulators based on similar structures such as silicon PCW MZI [14] and EO polymer filled silicon slot waveguide MZI [33], while our EO polymer filled silicon slot PCW MZI modulator combines the benefits from both the slow-light PCW [14] and the silicon-organic hybrid (SOH) structure [33]. In Reference [14], Nguyen, et al, demonstrated a silicon Mach-Zehnder modulators with 90μm-long lattice-shifted photonic crystal waveguides with $n_g$=20~30. By utilizing the plasma dispersion effect on p-n diode, digital modulation at a data rate of 40Gbit/s, optical bandwidth of 12.5nm were experimentally demonstrated. However, only a peak-to-peak driving voltage, $V_{pp}$=5.3V (instead of $V_\pi$), is reported in this reference, so we theoretically estimate the $V_\pi$ of the modulator [23, 53, 58] to be $V_\pi = \pi \big/ (\frac{2\pi}{\lambda} \Delta n_{eff} n_g L)$ $= \pi \big/ (\frac{2\pi}{1550nm} \cdot 1.6 \times 10^{-5} \cdot 30 \cdot 90 \mu m)$ =18V for apples-to-apples comparison. This large $V_\pi$ value leads to a large $V_\pi \times L$ product of 1.62V×mm at 10Gbit/s. In reference [33], Palmer, et al, reported a MZI modulator in which EO polymer is filled into a 1.5mm-long slot waveguide with slot width of 80nm. This modulator was demonstrated with in-device $r_{33}$ of 15pm/V, energy consumption of 320fJ/bit at 10Gbit/s digital modulation, and $V_\pi$=2.5V which corresponds to a $V_\pi \times L$=3.75V×mm. In a recent reference published by the same group [59], Alloatti, et al, demonstrated a MZI modulator with 500μm-long 120nm-wide-slot waveguide filled with EO polymer, in which they reported a 3dB bandwidth of 100GHz, but a relatively higher $V_\pi \times L$=11V×mm. In addition, recently, a rather specialized SOH modulator based on plasmonic waveguide has been demonstrated by Melikyan, et al, with bit rate of 40Gbit/s, energy consumption of 60fJ/bit, optical bandwidth of 120nm[45]. Considering the reported $V_\pi \times L$=1.3Vmm and device length of 29um, its value of $V_\pi$ is as large as 45V, which is prohibitive for many applications. With the continuous efforts and positive contributions made by researchers in this area, more research progress with improved device performance is expected to be reported in the future.

In our future work, the $V_\pi \times L$ product can also be further reduced by using more efficient organic EO materials, such as supramolecular organic EO glasses and binary EO polymers exhibiting intrinsic Pockels coefficients greater than 300 pm/V [60]. And also, the poling efficiency can be improved by using pyroelectric poling technique [61]. A transmission line can be designed to drive the modulator as a traveling wave device, in order to achieve modulation frequency over 100GHz [59, 62]. Recently, the SOH slot waveguide structures have been developed for digital modulations more advanced modulation formats such as QPSK and 16QAM with >1Tb/s [63, 64], thus, our future work will also include driving our modulator using high-speed digital signals [14, 19, 52] and also using this modulator for advanced modulation formats and coherent modulation scheme [63, 64]. In addition, potential stability of the modulator, a common issue for almost all polymer based devices, can be further improved by developing new EO polymers with higher glass transition temperatures and crosslinking chemistry, and by hermetically sealing the EO polymer and removing oxygen in the device packaging [65]. More detailed studies of EO polymer thermal stability indicate that operation up to 150℃ results in a change of EO coefficients < 10% [66, 67].

## 6. CONCLUSION

In summary, we demonstrate a broadband, low-power, low-dispersion and compact optical modulator based on a silicon slot PCW filled with EO polymer. Benefiting from the combined enhancement provided by the slow-light effect and high-$r_{33}$ polymer, the voltage-length product of the modulator is measured to be as small as $V_\pi \times L$=0.282V×mm, corresponding to a record-high effective in-device $r_{33}$ of 1230pm/V ever demonstrated. The silicon PCW is selectively doped to reduce the RC time delay and to achieve high-speed modulation. A backside gate technique is applied to our silicon PCW device to enhance device performance. Assisted by the backside gate voltage of 300V, the 3-dB bandwidth of the modulator is demonstrated to be 15GHz, and a modulation response up to 50GHz is observed. To the best of our knowledge, this is the first experimental demonstration of a 50GHz SOH PCW MZI modulator, although a 50GHz silicon /polymer hybrid slot MZI modulators [33], a 40GHz silicon /polymer hybrid PCW band-edge-shift modulator [42], and a 40Gbit/s monolithic silicon PCW MZI modulators [14] were previously demonstrated. In addition, the power consumption of the modulator is measured to be 24mW at 10GHz, and the estimated value of energy consumption per bit for a potential digital modulation is approximately 94.4fJ/bit at 10Gbit/s based on measured $V_\pi$ at 10GHz [26, 33, 45, 49, 50]. By using the band-engineered PCWs, the modulator is demonstrated to have a low-dispersion optical spectrum range as wide as 8nm, which is a factor of ~10X better than other modulators based on non-band-engineered PCWs [2, 19] and ring resonators [54] which have narrow operating optical bandwidth of <1 nm.

## ACKNOWLEDGEMENT

The authors would like to acknowledge the Air Force Research Laboratory (AFRL) for supporting this work under the Small Business Technology Transfer Research (STTR) program (Grant No. FA650-12-M-5131) monitored by Drs. Rob Nelson and Charles Lee.